\documentclass[10pt]{article}

\usepackage{amsfonts}
\usepackage{amsmath, graphicx, epsfig,psfrag}

\usepackage{caption}
\usepackage{subcaption}
\usepackage{multirow}
\usepackage{rotating}

\usepackage[usenames]{color}
\usepackage{epstopdf,url}

\usepackage{undertilde}

\numberwithin{equation}{section}

\begin{document}

\newcommand{\btheta}{\mbox{\boldmath$\theta$}}

\newcommand{\blambda}{\mbox{\boldmath$\lambda$}}
\newcommand{\bzeta}{\mbox{\boldmath$\zeta$}}

\newcommand{\bnu}{\mbox{\boldmath$\nu$}}
\newcommand{\bmu}{\mbox{\boldmath$\mu$}}
\newcommand{\bomega}{\mbox{\boldmath$\omega$}}
\newcommand{\bOmega}{\mbox{\boldmath$\Omega$}}
\newcommand{\bSigma}{\mbox{\boldmath$\Sigma$}}
\newcommand{\bGamma}{\mbox{\boldmath$\Gamma$}}

\newcommand{\ez}{\mathbb{E}}
\newcommand{\pz}{\mathbb{P}}
\newcommand{\convp}{\stackrel{p}{\longrightarrow}}
\newcommand{\nr}{n \rightarrow \infty}

\newcommand{\bthetasmall}{\mbox{\small \boldmath$\theta$}}

\newcommand{\bs}{}

\baselineskip20truept \pagestyle{plain}

\pagenumbering{arabic}

\title{Real time analysis of epidemic data}
\author{Jessica Welding and Peter Neal}
\maketitle

\begin{abstract}
Infectious diseases have severe health and economic consequences for society. It is important in controlling the spread of an emerging infectious disease to be able to both estimate the parameters of the underlying model and identify those individuals most at risk of infection in a timely manner. This requires having a mechanism to update inference on the model parameters and the progression of the disease as new data becomes available. However, Markov chain Monte Carlo (MCMC), the {\it gold standard} for statistical inference for infectious disease models, is not equipped to deal with this important problem. Motivated by the need to develop effective statistical tools for emerging diseases and using the 2001 UK Foot-and-Mouth disease outbreak as an exemplar, we introduce a Sequential Monte Carlo (SMC) algorithm to enable real-time analysis of epidemic outbreaks.
Naive application of SMC methods leads to significant particle degeneracy which are successfully overcome by particle perturbation and incorporating MCMC-within-SMC updates.
\end{abstract}

\section{Introduction} \label{S:Intro}

Markov chain Monte Carlo (MCMC) has been the leading tool is analysing infectious disease models over the last 20 years or so since the pioneering work of \cite{Gibson1997}, \cite{GibsonRenshaw1998} and \cite{OR99}. MCMC has become the {\it gold standard} for analysing epidemic models and for inferring the parameters of the models. The popularity of MCMC methods in analysing infectious disease data is largely due to the fact that epidemic data is almost always incomplete in the sense that, we typically know when individuals show symptoms to a disease, but not when the individuals became infected. Therefore the observed epidemic data does not typically admit a tractable likelihood and data augmentation techniques are required to infer the parameters of the model and the unobserved data (occult infections) which are often of interest in their own right.

MCMC has been used both for analysing epidemics in progress (\cite{OR99}) and completed epidemics with the majority of attention on the {\it post-hoc} analysis of completed epidemics. However, most practical interest in epidemic modelling is analysing the epidemic as it progresses to inform on actions such as control measures to limit the progress of the disease. Unfortunately MCMC methods are not conducive to estimating the parameters and state of the epidemic as it progresses with the performance of the MCMC algorithm becoming slower with poor mixing as more data becomes available. Therefore this paper seeks to explore an alternative to MCMC, namely SMC (Sequential Monte Carlo) methods which can be utilised to update the posterior distribution of the parameters and the state of the epidemic as the disease progresses.

Alternatives to MCMC for analysing infectious disease models, for example, ABC (approximate Bayesian computation), see, for example, \cite{Baguelin10} and \cite{KNP16} and emulation, see, for example, \cite{Andrianakis}. ABC has become a popular tool in analysing epidemic data since epidemic models are usually easy to simulate from. However, ABC based methods do not address the problem of updating estimates of the epidemic process as the disease progresses and new data becomes available.

Sequential Monte Carlo (SMC) methods, also known as particle filtering methods, \cite{SMCbook} are designed to update the posterior distribution of the parameters and the state of a stochastic process as it progresses. SMC algorithms use particles, samples from the posterior distribution of the parameters and the stochastic process of interest at time $t$ along with new data observed at time $t+1$ to obtain updated estimates of the posterior distribution of the parameters and the stochastic process of interest at time $t+1$. This process will form the building block for the SMC algorithm for epidemic models introduced in this paper.


Standard SMC algorithms cannot easily be applied to epidemic models since at each time point newly observed data, for example, those who have shown symptoms to the disease, will often not be consistent with the state of the epidemic in a given particle. This means to avoid significant particle degeneracy we need to adapt particles to be consistent with the newly observed data. It is difficult beyond the simplest homogeneously mixing epidemic model to adapt particles without introducing bias into the particles. To enable us to correct for the biasing of particles we employ MCMC within the SMC algorithm to update the particles. This allows us to exploit the considerable research into, and efficient algorithms for, MCMC for epidemic models within a framework which utilises the strength and speed of SMC methods for updating the posterior distribution of the parameters and the underlying state of the epidemic as new data becomes available.

The paper is structured as follows. In Section \ref{S:Overview} we address the question of real-time analysis of epidemics using the 2001 Foot-and-mouth disease (FMD) outbreak in the UK as a motivating example. We present an outline of the MCMC within SMC algorithm to be utilised in this paper. In Section \ref{S:Model}, we describe the discrete time epidemic model and along with the likelihood. In Section \ref{S:Algorithm}, we present full details of the MCMC-within-SMC algorithm. In Section \ref{S:Theory}, we provide a brief discussion of how the MCMC-within-SMC algorithm from a theoretical perspective.
 In Section \ref{S:Simulation}, we explore the performance of the MCMC-within-SMC algorithm through a simulation study before presenting in Section \ref{S:FMD} an analysis of the 2001 FMD outbreak in the county of Cumbria, the most affected county in the UK by the 2001 FMD outbreak. Finally, in Section \ref{S:Conc} we make a few concluding remarks identifying avenues for future work. 

\section{Overview of real-time analysis for epidemic models} \label{S:Overview}

In this Section, we address the key question of, what does real-time analysis of epidemics mean? In particular, we put the question into the context of the data, model and inference algorithm, to motivate the analytical procedure that we present in Section \ref{S:Algorithm} and the analysis of the 2001 Foot-and-Mouth disease outbreak presented in Section \ref{S:FMD}. We start with a description of the  2001 Foot-and-Mouth disease outbreak in Section \ref{ss:FMD:desc} outlining why it is the motivating example for our work. In Section \ref{ss:Overview:daily}, we explore the meaning of real-time analysis for epidemic models. We define real-time analysis of data to be on a  suitable time-frame as the data becomes available, so that an interested party can act in a meaningful manner to
affect the underlying stochastic process, in our case the epidemic, on the basis of the analysis of the
data undertaken. In Section \ref{ss:Overview:SMC}, we overview the sequential Monte Carlo (SMC) methodology developed in Section \ref{S:Algorithm}, highlighting the key considerations within the epidemic modeling framework.

\subsection{The 2001 UK Foot-and-Mouth disease outbreak} \label{ss:FMD:desc}

The main motivating example for this work is the 2001 Foot-and-Mouth disease (FMD) outbreak in the UK. This saw a major FMD epidemic take place between February and September 2001 with 2026 confirmed infected farms and a further 8585 farms culled as being considered {\it a priori} at high risk of being infected. This led to the killing of over 10 million cattle and sheep across the UK. Therefore the outbreak had a major impact on the UK economy with the National Audit Office estimating the cost to be over \pounds 3 billion to the public sector and \pounds 5 billion to the private sector, see \cite{NAO}. The epidemic originated in Essex in the South East of England but rapidly spread across Great Britain with the worst affected area being Cumbria in the North West of England was most severely affected with a total of 893 confirmed infectious farms. It is the outbreak in Cumbria, following \cite{Jewell09} and \cite{XN14}, which is the interest of our analysis.

The significance of the FMD outbreak along with the rich data available on when the disease was detected on a farm (notification date) and when a farm was culled (removal) date has led to substantial analysis of the data. Initial {\it ad-hoc} analysis of the data took place whilst the FMD epidemic outbreak was in progress   \cite{keeling2001dynamics}, \cite{ferguson2001foot} and \cite{ferguson2001transmission}. These initial findings found that cattle were both more infectious and susceptible to FMD than sheep but that this is balanced by the fact the number of sheep are greater. Larger farms, and in particular, fragmented farms were found to be more infectious and susceptible than smaller farms although this relationship is found to be non-linear in farm size (\cite{ferguson2001transmission}). The rapid transmission of FMD supported culling rather than vaccination in controlling the spread of the disease, \cite{keeling2001dynamics}, \cite{ferguson2001foot}. The general findings of the {\it in progress} analysis of FMD have largely be confirmed by {\it post-hoc} analysis of the disease. Much of the {\it post-hoc} statistical analysis of the FMD outbreak centres around the Cambridge-Edinburgh model (\cite{keeling2005models}) or variants of the model. These models model the transmission of FMD at the level of farms (base unit) with covariates, namely, the total number of sheep and cattle, defining the susceptibility and infectiousness of farms with a spatial kernel defining the interaction, and hence, the transmitability of FMD between farms. Examples of FMD analysis based on the  Cambridge-Edinburgh model include   \cite{kypraios2007efficient}, \cite{Jewell09}, \cite{deardon2010inference} and \cite{XN14}. \cite{deardon2010inference} used a discrete-time daily model whilst the other papers use a continuous time model and with all these papers applying computationally intensive MCMC to estimate the model parameters. In particular, the dates on which farms become infected and infectious are not observed and thus need to be imputed within the modeling framework. Depending on the model being infected and becoming infectious may or may not coincide. The {\it post-hoc} papers generally treat the infection times as unknown to be inferred using data augmentation as part of the MCMC algorithm whilst \cite{keeling2001dynamics} takes fixed length exposed (latent) and infectious periods for FMD to circumvent this problem. Overall this means that the approaches taken in the {\it post-hoc} analyses are not readily applicable to in progress analysis of a new FMD or similar disease outbreak.

Our approach is to take the 2001 FMD outbreak as a motivating example to develop statistical inferential techniques for epidemics in progress which are comparable in performance to the {\it gold standard} MCMC methods developed in, for example,  \cite{kypraios2007efficient}, \cite{Jewell09}, \cite{deardon2010inference} and \cite{XN14}. The statistical approaches developed in this paper can then be applied to a new epidemic outbreak to give robust understanding of the disease parameters and progression. This can be used to inform the implementation of control measures and actions without making the limiting assumptions necessary for the analysis in  \cite{keeling2001dynamics}, \cite{ferguson2001foot} and \cite{ferguson2001transmission}.

\subsection{Real-time analysis for epidemic models} \label{ss:Overview:daily}

For an emerging disease such as FMD in a large susceptible population efficient control of the disease is the major public health aim. In order to devise control strategies such as culling of farms in the cases of FMD or targeted vaccination for a range of human and animal diseases it is vital to be able to; estimate the parameters underlying the model for the disease, determine the probability that an individual (or other unit of interest such as a farm or household) is already infected and to identify those most at risk of infection in the short-to-medium term; on a time-scale which allows appropriate action to take place. Therefore assuming daily data which mainly arrives during a nominal {\it working} day of 9am to 6pm, say, we want to be able to analyse the data overnight (approximately 12 hours) and report back the key quantities of interest, so that action can be taken on the basis of the findings during the next {\it working} day. By contrast for the 2014 Ebola outbreak in West Africa case data is available from the World Health Organisation (WHO), \cite{WHOEbola} on a weekly basis. For  most epidemics processes, including FMD and Ebola,  the observed data represents only partial information on the epidemic process. As noted above the observed data for FMD consists of the date on which the farm becomes a notified premise, that is, symptoms are detected on the farm and the date on which a farm is culled (removed). There is no information about the day on which farms are infected although this information is crucial in writing down a tractable likelihood for statistical inference.

%
%

Given that the data we are analysing is assumed to be reported on a daily basis, we follow \cite{deardon2010inference} in  using a discrete time model for the data with full details of the model construction given in Section \ref{S:Model}. The choice of a discrete time model over a continuous time model is largely pragmatic to assist with the sequential Monte Carlo (SMC) algorithm employed to analyse the data on a day-by-day basis. The popularity of continuous time models for epidemics is largely an artifact of the early mathematical analysis of deterministic and stochastic epidemics via ODEs and Markov processes, respectively. Whilst the spread of a disease is a process taking place in continuous time, continuous time models usually assume a constant rate of contact (infection) between individuals throughout the course of the day which does not take into account that interactions between individuals are different at different times of the day. In \cite{Neal16} a simple household model which incorporated time of day effects, {\it i.e.}~different interactions between individuals during the day and the night, was studied. It can be shown using the approach taken in  \cite{Neal16} that if the mean length of the infectious period is 3 or more days long there is little difference in the probability of a major epidemic between two models with the same mean number of contacts per day, a model which incorporates time of day effects and the standard continuous time model. Similarly, the discrete time model which condenses the interaction between individuals into a daily probability of infection offers a reasonable approximation of the epidemic process provided that individuals are typically infectious for a number of days. We assume that the underlying epidemic dynamics are $S \rightarrow I \rightarrow N \rightarrow  R$ with individuals starting off susceptible, on becoming infected an individual becomes immediately infectious the next day. After a given (random) period of time an infective displays symptoms and becomes a notified case. This could result in immediate removal of the individual or removal/recovery of an individual may occur some days later. In the case where notification date corresponds to removal date the model simplifies to an $S \rightarrow I \rightarrow R$ epidemic model. The inclusion of an exposed state (latent period) between being infected and becoming infectious can readily be incorporated into the model.

%
%

\subsection{Sequential Monte Carlo (SMC) methods for epidemic models} \label{ss:Overview:SMC}

We turn to the analytical procedure which will be presented in detail in Section \ref{S:Algorithm}. Given that the data are assumed to be observed daily, we employ an SMC algorithm (particle filter) to estimate the parameters and the occult infections (individuals infected, but not yet detected), \cite{Jewell09}. That is, we use data augmentation of key unobserved events in the epidemic process to assist the analysis. In order to outline the process we represent the parameters of the model by $\btheta$, the observed data on day $t$ by $x_t$ with $\mathbf{x}_{0:t} = (x_0, x_1, \ldots, x_t)$ (we denote the day on which the first case is detected as day 0) and the augmented data on day $t$ by $y_t$. Let $\tau <0$ denote the (unknown) day upon which the disease was introduced into the population with  $\mathbf{y}_{\tau:t} = (y_\tau, y_{\tau+1}, \ldots, y_t)$. Then on day $t \geq 0$, we are interested in samples $(\btheta, \mathbf{y}_{\tau:t})$ from $\pi ( \btheta, \mathbf{y}_{\tau:t} |\mathbf{x}_{0:t})$. Moreover, on day $t+1$ with additional data $x_{t+1}$, we want to utilise our samples from the posterior distribution on day $t$ to inform our draws from the posterior distribution on day $t+1$ without reanalysing the entire data from scratch. This is the motivation behind employing an SMC algorithm.


The SMC methodology has successfully been applied to a range of problems requiring rapid {\it online} analysis of data, such as target tracking, \cite{Nemeth14}, and data streaming, \cite{Zhu17}. The epidemic timescale is sedate by comparison but is sufficiently fast paced that the re-evaluation of the data on a daily basis using MCMC is impractical for moderate-to-large data sets. This relative slow pace enables us to incorporate elements of MCMC into our analysis to counter the common problem of particle degeneracy within SMC and also a problem specific issue of the augmented data  $\mathbf{y}_{\tau:t} $ often not being compatible with the new data $x_{t+1}$. That is, there often exists at least one newly detected case at time $t+1$ which does not have an infection time in $\mathbf{y}_{\tau:t} $. Furthermore we explore using MCMC to {\it seed} the initial particles for the SMC. An outline of the process is as follows.

We select $T>0$, as an initial time point to analyse the data, we obtain samples from $\pi ( \btheta, \mathbf{y}_{\tau:T} |\mathbf{x}_{0:T})$ using MCMC. The MCMC is run for $B+M\times N$ iterations, where $B$ is a burn-in period, $N$ is the number of particles to be used in the SMC algorithm and $M$ is a thinning parameter with every $M^{th}$ realisation from the MCMC output after the burn-in used to form a particle $(\btheta, \mathbf{y}_{\tau:T})$ with each particle given equal weight, $w$, nominally 1.
Alternatively, multiple MCMC runs can be used each with burn-in $B$ to generate $N$ particles with which to initiate the SMC. Then for $t \geq T$:
\begin{enumerate}
\item Let $\{ (\btheta^{i}_t, \mathbf{y}_{\tau:t}^{i}); 1 \leq i \leq N \}$ denote samples from $\pi ( \btheta, \mathbf{y}_{\tau:t} |\mathbf{x}_{0:t})$ and let $w_t^i$ denote the weight associated with particle $i$.
\item For $i=1,2,\ldots, N$, update $\mathbf{y}_{\tau:t}^{i}$ to be consistent with the new data $x_{t+1}$ and update the weight $w_t^i$ accordingly. This ensures every detected case at time $t+1$ has an infection time prior to time $t+1$.
\item Sample $N$ particles $\{ (\tilde{\btheta}^{j}_t, \tilde{\mathbf{y}}_{\tau:t}^{j}); 1 \leq j \leq N \}$ with replacement from  $\{ (\btheta^{i}_t, \mathbf{y}_{\tau:t}^{i}); 1 \leq i \leq N \}$ with
\begin{eqnarray} \label{eq:over:1}
P \left( (\tilde{\btheta}^{j}_t, \tilde{\mathbf{y}}_{\tau:t}^{j}) = (\btheta^{i}_t, \mathbf{y}_{\tau:t}^{i})\right) = \frac{w_t^i}{\sum_{l=1}^N w_t^l}.
\end{eqnarray}
\item For $i=1,2,\ldots, N$, in parallel:
\begin{enumerate}
\item Sample $y_{t+1}$ (new infections at time $t$) from
\begin{eqnarray} \label{eq:over:2}
\pi (\mathbf{y}_{t+1} | \tilde{\btheta}^{i}_t, \tilde{\mathbf{y}}_{\tau:t}^{i}, \mathbf{x}_{0:t+1})
\end{eqnarray}
and set $(\btheta^i_{t+1}, \mathbf{y}^i_{\tau:t+1}) = (\tilde{\btheta}^i_t, (\tilde{\mathbf{y}}^i_{\tau:t}, y^i_{t+1}))$.
\item Starting with $(\btheta^i_{t+1}, \mathbf{y}^i_{\tau:t+1})$ generated in step a), use $n_p$ iterations of MCMC to update the parameter and augmented data.
\item The final value of $(\btheta^i_{t+1}, \mathbf{y}^i_{\tau:t+1})$ from the MCMC gives the $i^{th}$ particle to take forward to the next time-point. The associated weight for the particle is $w_{t+1}^i = \pi ( \btheta^i_{t+1}, \mathbf{y}^i_{\tau:t+1}| \mathbf{x}_{0:t+1})$ and it suffices that this is only known up to a constant of proportionality.
\end{enumerate}
\end{enumerate}
The details of the steps are provided in Section \ref{S:Algorithm}. There are two key points to address. Firstly, how do we adapt the data $\mathbf{y}_{\tau:t}^{i}$ (and weighting $w^i_t$) to be consistent with $x_{t+1}$, whilst ensuring that the samples are still from the correct posterior distribution? The procedure we use ensures this for homogeneously mixing epidemics but more generally the adaption step will lead to a small bias. This is corrected for by the MCMC step which targets the correct posterior distribution. This leads onto the second question, what MCMC updating schema to use and how large should $n_p$ be? There is a wealth of knowledge for MCMC epidemic models, see, for example, \cite{GibsonRenshaw1998}, \cite{OR99}, \cite{Jewell09} and \cite{XN14}, which can be utilised to devise the updating schema. The value of $n_p$ is interesting and we present a partial answer in Section \ref{S:Theory} with further exploration of $n_p$ presented elsewhere. Letting $n_p \rightarrow \infty$, we are guaranteed that each particle will be a draw from the posterior distribution of interest using standard properties of Markov chain convergence. However, for practical purposes we require $n_p$ to be relatively small as the MCMC step is the time consuming component of the algorithm and we are seeking to present a viable alternative to large scale MCMC. We observe that the MCMC samples start with approximate draws from $\pi (\cdot | \mathbf{x}_{0:t+1})$ assuming that the marginal densities $\pi (\btheta, \mathbf{y}_{\tau:t} | \mathbf{x}_{0:t})$ and  $\pi (\btheta, \mathbf{y}_{\tau:t} | \mathbf{x}_{0:t+1})$ are similar. Also, we only require $n_p$ to be large enough such that $\{ (\btheta^i_{t+1}, \mathbf{y}^i_{\tau:t+1}); 1 \leq i \leq N \}$ is a representative sample from $\pi (\cdot | \mathbf{x}_{0:t+1})$ and not for greater mixing within each MCMC runs. 

%

%
%

The final observation before studying the model and algorithm in more detail is the question of computational cost. Throughout the computationally expensive elements of the algorithms are the MCMC iterations with all other computations taking insignificant amounts of time in comparison. Therefore a naive comparison of cost would be the total number of MCMC iterations required at each time point. This does not reflect the true cost to the practitioner who is likely even with a standard PC to have multiple processors available and thus able to exploit the embarrassingly parallel nature of the particle updates with $n_p$ updates of $N$ particles in practice being many times faster than $N \times n_p$ updates within a single MCMC chain.


\section{Model and likelihood} \label{S:Model}

In this Section we outline the generic model to be analysed and construct the likelihood. The model is constructed with analysis of the 2001 FMD outbreak in mind but is more widely applicable.

We assume that the population is closed and of size $n$ with the individuals labelled $i=1,2,\ldots, n$. There is assumed to be one initial infective, denoted $\nu$, who is responsible for introducing the disease to the population with all other infections via infectious transmissions within the population. We consider an
\begin{eqnarray} \label{eq:model:1}
\mbox{{\bf S}(usceptible) } \rightarrow \mbox{{\bf I}(nfective)} \rightarrow \mbox{{\bf N}(otified)} \rightarrow \mbox{{\bf R}(emoved)}
\end{eqnarray}
epidemic model with the special case where notification and removal occur instantaneously being the $SIR$ epidemic model. The extension to $SEIR$ epidemic models is straightforward.
As noted in Section \ref{S:Overview}, we assume a discrete time model for the disease transmission with time $t \in \mathbb{Z}$ with reference point day 0 corresponding to the date of the first notified case of the disease. 

On a given day each individual belongs to one of the four categories; susceptible, infectious, notified or removed. For $t \in \mathbb{Z}$, let $\mathcal{S}_t, \mathcal{I}_t, \mathcal{N}_t$ and $\mathcal{R}_t$ denotes the set of individuals who are susceptible, infectious, notified and removed, respectively, at time $t$. On day $t$ an infective $i \in \mathcal{I}_t$ has probability $p_{ij}$ of making an infectious contact with an individual $j$, whereas a notified individual $k \in \mathcal{N}_t$ has probability $\kappa p_{kj}$ of making an infectious contact with an individual $j$ where $\kappa$ denotes the relative infectiousness of notified individuals to infectives. (Note that if $\kappa =0$ the model is indistinguishable from an $SIR$ epidemic model.) If at least one infectious contact is made with a susceptible individual $j$ on day $t$, then individual $j$ will become infected on day $t+1$ and start to make infectious contacts. For $1 \leq i,j \leq N$, the probability $p_{ij}$ will depend upon (infection) parameters, $\btheta$, and covariate information $\mathbf{z}_i$ and $\mathbf{z}_j$ for individuals $i$ and $j$, which we denote $h (\btheta, \mathbf{z}_i, \mathbf{z}_j)$. The simplest choice of $h (\btheta, \mathbf{z}_i, \mathbf{z}_j)$ is the homogeneously mixing epidemic model with $h (\btheta, \mathbf{z}_i, \mathbf{z}_j) = (1-p)$, where $p$ is the (avoidance) probability of avoiding infection from a given infective. In Section \ref{S:Simulation} for the simulation study we consider a spatial model setting $h (\btheta, \mathbf{z}_i, \mathbf{z}_j) = (1-p) \exp(- \gamma d(i,j))$, where $d (i,j)$ denotes the (Euclidean) distance between individuals $i$ and $j$. This model is further developed to include covariates such as farm size in Section \ref{S:FMD} for the FMD outbreak. We could also allow $h (\cdot, \cdot, \cdot)$ to be a function of time but do not consider that extension in this paper. We assume that an individual $i$ infected on day $t$, say, is infectious for days $t + 1, \ldots, t + Q_i$ before becoming a notified case for days $t+Q_i +1, \ldots t + Q_i +U_i$ and then removed from day $t + Q_i +U_i +1$ onwards. The infectious period distributions $Q_i$'s are assumed to be independent and identically distributed according to an arbitrary, but specified, integer valued distribution, $Q$. Let $g_Q (\cdot; \btheta)$ denote the probability mass function of $Q$ which we allow to depend upon the parameters of the model. That is, we assume that the distributional family  to which $Q$ belongs is known, but not necessarily its parameters. Since the notification and removal dates are assumed to be observed we do not explicitly model the distribution of the $U_i$'s. The epidemic ceases once there are no more infectives or notified individuals in the population.

Let $\tau (<0)$ denote the day upon which the original infective, $\kappa$, becomes infected and note that both $\kappa$ and $\tau$ are assumed to be unknown. Returning to notation of Section \ref{S:Overview}, we take $\mathbf{x}_{0:t}$ and $\mathbf{y}_{\tau:t}$ to denote the observed and unobserved data, respectively, pertaining to individuals infected up to and including day $t$. Now $\mathbf{x}_{0:t} = (\mathbf{n}_{0:t}^O, \mathbf{r}_{0:t})$, the notification times ($\mathbf{n}_{0:t}^O$) of individuals notified up to and including day $t$ and the corresponding removal times ($\mathbf{r}_{0:t}$) if these occur on day $t$ or before. Whilst, $\mathbf{y}_{\tau:t} = (
\mathbf{i}_{\tau:t}^O, \mathbf{i}_{\tau:t}^U, \mathbf{n}_{\tau:t}^U)$ the infection times ($\mathbf{i}_{\tau:t}^O$) of individuals notified  up to and including day $t$, the infection times ($\mathbf{i}_{\tau:t}^U$) of occult individuals on day $t$ (individuals infected on day $t$ or before but whom do not become notified individuals until after day $t$) and the notification times ($\mathbf{n}_{\tau:t}^U$) of occult individuals on day $t$. The $\mathbf{n}_{\tau:t}^U$ denote the time of future events which assist in constructing a tractable likelihood. There is a one-to-one relationship between $\{ (\mathcal{S}_s, \mathcal{I}_s, \mathcal{N}_s,\mathcal{R}_s); \tau \leq s \leq t \}$ and $(\mathbf{x}_{0:t}, \mathbf{i}_{\tau:t}^O, \mathbf{i}_{\tau:t}^U)$, and we can use the representations of the data interchangeably.

Given $\btheta$, we have that
\begin{eqnarray} \label{eq:model:2}
\pi (\mathbf{x}_{0:t}, \mathbf{y}_{\tau:t} | \btheta) &=& \prod_{s=\tau}^{t-1} \left\{ \prod_{l \in \mathcal{S}_{s+1}} P_s (l; \btheta)  \prod_{l \in \mathcal{S}_s \backslash \mathcal{S}_{s+1}} \{1 -P_s (l; \btheta) \} \right\} \nonumber \\
&& \times \prod_{k \not\in \mathcal{S}_t} g_Q (n_k - i_k),
\end{eqnarray}
where $i_k$ and $n_k$ denote the infection and notification time of individual $k$, respectively, and $P_s (l; \btheta)$ is the probability individual $l$ avoids infection on day $s$ which is given by
\begin{eqnarray} \label{eq:model:3}
P_s (l; \btheta) = \prod_{k \in \mathcal{I}_s} (1- p_{kl}) \times \prod_{k \in \mathcal{N}_s} (1-\kappa p_{kl}).
\end{eqnarray}
From \eqref{eq:model:2} with an appropriate prior on $\btheta$, it is straightforward to obtain the posterior density, $\pi ( \mathbf{y}_{\tau:t}, \btheta |\mathbf{x}_{0:t})$ up to a constant of proportionality with
\begin{eqnarray} \label{eq:model:4}
\pi ( \mathbf{y}_{\tau:t}, \btheta |\mathbf{x}_{0:t}) \propto \pi (\mathbf{x}_{0:t}, \mathbf{y}_{\tau:t} | \btheta) \times \pi (\btheta). \end{eqnarray}
The equation \eqref{eq:model:4} will play a key role in the construction of the SMC algorithm in Section \ref{S:Algorithm}. Note that in contrast to {\it post-hoc} analysis of epidemics where the key interest is in the marginal density $\pi (\btheta |\mathbf{x}_{0:t})$, we are also interested $\mathbf{y}_{\tau:t}$, both for the sequential updating of parameters and predictions of the future course of the epidemic process.

\section{Algorithm} \label{S:Algorithm}

This Section is at the core of a procedure for the real-time analysis of epidemics. In Section \ref{ss:MCMC} we outline the MCMC step which is utilised in both the initialisation of the particles for the SMC on day $T$ and updating the particles from one day to the next. In Section \ref{ss:Initialisation} we discuss using MCMC to initialise the particles for the SMC. The main focus of this Section is the details of the SMC algorithm in Section \ref{ss:SMC}. Here we highlight two key novelties in our procedure, the modifying of particles to be consistent with the new observed data (Section \ref{sss:consistent}) and the use of short MCMC runs to update the particles and guard against particle degeneracy (Section \ref{sss:jitter}).

\subsection{MCMC step} \label{ss:MCMC}

We outline the MCMC step which forms the bedrock of the SMC algorithm. Most of the features incorporated within the MCMC step to make it effective are based upon the extensive study of MCMC algorithms for epidemic models although we introduce novelty in the updating of the number of occult cases.

In the MCMC step we need to update three components; the model parameters, $\btheta$, the infection times of the notified cases, $\mathbf{i}_{\tau:t}^O$ and the total number, as well as the infection and notification times, of the occult cases, $(\mathbf{i}_{\tau:t}^U, \mathbf{n}_{\tau:t}^U)$. We update the three components in turn.

\subsubsection{Step 1: Updating $\btheta$}   \label{sss:theta}

Let $\btheta = (\blambda, \bzeta)$, where $\blambda$ and $\bzeta$ denote the parameters underpinning the infectious process and the infectious period distribution, $Q$, respectively. Then provided that independent priors are chosen for $\btheta$ and $\bzeta$, {\it ie.}~$\pi (\btheta) = \pi (\blambda) \pi (\bzeta)$, we have from \eqref{eq:model:2} and \eqref{eq:model:4} that
\begin{eqnarray} \label{eq:MCMC:1}
\pi (\btheta | \mathbf{x}_{0:t}, \mathbf{y}_{\tau:t}) = \pi (\blambda | \mathbf{x}_{0:t}, \mathbf{y}_{\tau:t}) \times \pi (\bzeta | \mathbf{x}_{0:t}, \mathbf{y}_{\tau:t}).
\end{eqnarray}
Therefore we update $\blambda$ and $\bzeta$ separately as with continuous time models, see for example, \cite{OR99} and \cite{Jewell09}.

For $\blambda$ we use random walk Metropolis to update the parameter with proposal covariance matrix $\Sigma_\lambda$. The optimal choice of $\Sigma_\lambda$ should result in acceptance rate of approximately $25\%$ and resemble the correlation structure in the parameters, see \cite{RobRos01}, Section 7. In the initial MCMC to initialise the particles we tune  $\Sigma_\lambda$ adaptively starting from a scalar multiple of the identity matrix, whereas for update steps with the SMC algorithm we utilise the sample from the posterior distribution at the previous time point to inform the choice of  $\Sigma_\lambda$, see Section \ref{sss:jitter}.

%
%

For $\bzeta$, the update of the parameters will be more distribution specific as often it will be the case that for some of the components of $\bzeta$ Gibbs sampling steps can be used. For the other components we again use random walk Metropolis.

\subsubsection{Step 2: Updating $\mathbf{i}_{\tau:t}^O$}   \label{sss:known}

For the updating of $\mathbf{i}_{\tau:t}^O$, we employ an independence sampler similar to that used in \cite{XN14,LN16} for updating the infection times relative to the removal times. We choose a random sample $\mathcal{F}$, of size $m$, of individuals from those in $\mathcal{N}_t \cup \mathcal{R}_t$. For each individual $l \in \mathcal{F}$, we draw a new infectious period distribution $q_l$ from $g_Q (\cdot)$ and set $i_l^\prime = n_l - q_l$, the proposed new infection time of individual $l$. For all individuals not in $\mathcal{F}$, the infection times remain unchanged. We then compute the acceptance probability for the proposed move, noting that the proposal distribution is chosen to lead to a cancellation with the infectious period terms in \eqref{eq:model:2}, {\it cf.}~\cite{XN14}. It is shown in \cite{LN16} that it is optimal for $m$ to be chosen such that approximately $25\%$ of iterations are accepted. Therefore we monitor the acceptance rate and adjust $m$ accordingly with $m$ generally increasing with the number of observed notifications. This step can lead to $\nu$ and/or $\tau$ being updated.


\subsubsection{Step 3: Updating $(\mathbf{i}_{\tau:t}^U, \mathbf{n}_{\tau:t}^U)$}   \label{sss:occult}

For the occult individuals there are two types of changes; either we change the number of occult infections or we change the times of the existing occult infections. In each iteration we perform both changes.

For updating the infection times of the occult cases the procedure is very similar to step 2. We choose a random sample $\mathcal{F}$, of size $m_U$, of occult individuals. For each individual $l \in \mathcal{F}$, we draw a new infectious period distribution $q_l$ from $g_Q (\cdot)$. However since the notification time of the occult individuals is not fixed, we also draw $h_l$ uniformly at random from $\{0,1,\ldots, q_l-1\}$ and set $i_l^\prime = t-h_l$ and $n_l^\prime = i_l^\prime+q_l$ giving both a new infection and notification time for individual $l$. Again for all individuals not in $\mathcal{F}$, the infection and notification times remain unchanged and we compute the acceptance probability for the proposed move. As above we monitor and adjust $m_U$ so that approximately $25\%$ proposed moves are accepted.

For changing the number of occult infections, we set a maximum change in the number of occult cases $e_u$ and draw $c$, the change in the number of occult infections, uniformly at random from $\{-e_u, - (e_u-1), \ldots, -1, 1, 2, \ldots, e_u \}$. If $c <0$, we randomly select $-c$ occult individuals and propose they become susceptible, conditional upon there being at least $-c$ occult individuals in the population. Whilst, if $c>0$, we randomly select $c$ susceptible individuals and propose they become occult individuals, conditional upon there being at least $c$ susceptible individuals in the population. For each proposed new occult individual, $l$, we draw an infectious period $q_l$ from $g_Q (\cdot)$ and $h_l$ uniformly at random from $\{0,1,\ldots, q_l-1\}$ to set $i_l^\prime = t-h_l$ and $n_l^\prime = i_l^\prime+q_l$ as above. It is then straightforward to compute the acceptance probability for the proposed move using \eqref{eq:model:2}.

\subsection{MCMC initialisation} \label{ss:Initialisation}

To generate the initial particles for the SMC algorithm at time $T$, we run an MCMC algorithm for $B+NM$ iterations keeping the output from every $M^{th}$ iteration of the MCMC algorithm after the $B$ burn-in iterations. We utilise the MCMC steps outlined in Section \ref{ss:MCMC} with arbitrary initial parameter values. A small number of occult cases are assigned to the population and the initial augmented data (infection and notification) times are then simulated using the infectious period distribution $Q$. The augmented data is then adjusted as necessary to ensure that it is consistent with the observed data and leads to a valid realisation of the epidemic process. This procedure leads to the MCMC algorithm quickly finding the posterior distribution provided that reasonable parameter values are chosen with an appropriate choice of $\Sigma_\lambda$ for the random walk Metropolis updates of the parameters.

We employ an adaptive RWM algorithm similar to \cite{Haario01}, Section 2, with adaption restricted to the burn-in period.  We start with $\Sigma_\lambda$ to be a multiple $\alpha$ of the identity matrix. During the burn-in period we adapt $\alpha$ by increasing (decreasing) it when a proposed move is accepted (rejected) using the updating schema in \cite{XN14}, (12) and (13) which leads to a long term acceptance rate of approximately $25\%$. At the end of the burn-in period we fix $\Sigma_\lambda$ to be
\begin{eqnarray} \label{eq:Sigma:1}
\Sigma_\lambda = (1- \xi) s_d \Sigma_B + \xi \tilde{\alpha} s_d I_d, \end{eqnarray} where $\Sigma_B$ denotes the empirical covariance matrix for the parameters at the end of the burn-in period, $I_d$ is the $d$ dimensional identity matrix and $s_d = 2.38^2/d$ is the optimal scaling parameter for RWM, see \cite{RGG97}. We take $\xi =0.05$ and $\tilde{\alpha}=0.1$. The inclusion of $\xi \tilde{\alpha} s_d I_d$ in \eqref{eq:Sigma:1} is a safety measure to avoid problematic (singular) values of $\Sigma_B$, see \cite{RobRos09}.

In the early stages of the epidemic with $T$ small and a few infectives the evaluation of the likelihood is quick. Also the mixing of the parameters and augmented data is generally better in the early stages of the epidemic when there is little data. Therefore it is usually practical to use MCMC to initialise the SMC with for example, $B=10,000$, $N=1,000$ and $M=50$ requiring $60,000$ iterations.

\subsection{Sequential Monte Carlo} \label{ss:SMC}

For each $t >T$, we seek to utilise our sample $\{ (\btheta^i_{t-1}, \mathbf{y}_{\tau:t-1}^i); i=1,2,\ldots, N\}$ from $\pi (\cdot | \mathbf{x}_{0:t-1})$ to generate a sample $\{ (\btheta^i_t, \mathbf{y}_{\tau:t}^i); i=1,2,\ldots, N\}$ from $\pi (\cdot | \mathbf{x}_{0:t})$. The first step is to use the marginal $\{ (\btheta^i_{t-1}, \mathbf{i}_{\tau:t-1}^i); i=1,2,\ldots, N\}$, where  $\mathbf{i}_{\tau:t-1}^i$ denotes the full set of infection times. That is, we marginalise over the unobserved (future) notification times of the occult individuals, since these are only used to assist with the MCMC and it gives us greater flexibility in matching the augmented data up to time $t-1$ with the new notification and removal times, $x_t$, on day $t$.

The next step is to consider the relationship between  $\pi (\cdot | \mathbf{x}_{0:t-1})$ and  $\pi (\cdot | \mathbf{x}_{0:t})$. We note that the new infections and notifications on day $t$ are independent, given $(\mathbf{x}_{0:t-1}, \mathbf{i}_{\tau:t-1})$, and thus
\begin{eqnarray} \label{eq:SMC:1}
&&\pi (\mathbf{x}_t, \mathbf{i}_t | \btheta, \mathbf{x}_{0:t-1}, \mathbf{i}_{\tau:t-1}) \nonumber \\&& \hspace*{1cm} = \pi (\mathbf{x}_t| \btheta, \mathbf{x}_{0:t-1}, \mathbf{i}_{\tau:t-1}) \times \pi (\mathbf{i}_t | \btheta, \mathbf{x}_{0:t-1}, \mathbf{i}_{\tau:t-1}). \end{eqnarray}
Therefore we can write
\begin{eqnarray} \label{eq:SMC:2}
\pi (\mathbf{i}_{\tau:t}, \btheta | \mathbf{x}_{0:t}) &=& \frac{ \pi (\mathbf{x}_t, \mathbf{i}_t | \btheta, \mathbf{x}_{0:t-1}, \mathbf{i}_{\tau:t-1}) \pi ( \mathbf{x}_{0:t-1}, \mathbf{i}_{\tau:t-1} | \btheta) \pi (\btheta)}{
\pi (x_t | \mathbf{x}_{0:t-1}) \pi (\mathbf{x}_{0:t-1})} \nonumber \\
&=& \frac{\pi (\mathbf{x}_t| \btheta, \mathbf{x}_{0:t-1}, \mathbf{i}_{\tau:t-1}) \times \pi (\mathbf{i}_t | \btheta, \mathbf{x}_{0:t-1}, \mathbf{i}_{\tau:t-1})}{\pi (x_t | \mathbf{x}_{0:t-1})} \nonumber \\ && \hspace*{3cm} \times \pi (\mathbf{i}_{\tau:t-1}, \btheta | \mathbf{x}_{0:t-1}) \nonumber \\
& \propto & \pi (\mathbf{x}_t| \btheta, \mathbf{x}_{0:t-1}, \mathbf{i}_{\tau:t-1}) \times \pi (\mathbf{i}_t | \btheta, \mathbf{x}_{0:t-1}, \mathbf{i}_{\tau:t-1}) \nonumber \\ && \hspace*{3cm} \times \pi (\mathbf{i}_{\tau:t-1}, \btheta | \mathbf{x}_{0:t-1}). \end{eqnarray}
Thus we need to take account of the two terms on the righthand side of \eqref{eq:SMC:1} in taking forward the particles sampled at time $t-1$.

%
%
%
%

\subsubsection{Consistent Particles} \label{sss:consistent}

The first observation from \eqref{eq:SMC:2} is that \\ $\pi (\mathbf{x}_t| \btheta, \mathbf{x}_{0:t-1}, \mathbf{i}_{\tau:t-1})$ will be 0 if there is at least one new notified case at time $t$ which does not have an infection time prior to time $t-1$. This can lead to substantial particle degeneracy with very few, or even no, particles having non-zero weight. The solution is to adjust $\mathbf{i}_{\tau:t-1}$ such that
\begin{eqnarray} \label{eq:SMC:3} \pi (\mathbf{x}_t| \btheta, \mathbf{x}_{0:t-1}, \mathbf{i}_{\tau:t-1}) \neq 0, \end{eqnarray} and our approach works so long as the number of new notifications at time $t$ is less than or equal to the total number of occult infectives at time $t-1$. The second observation from \eqref{eq:SMC:2}, before detailing the adjustment of the particles, is that $\mathbf{i}_t$ and $\mathbf{x}_t$ are conditionally independent given $\btheta$, $\mathbf{x}_{0:t-1}$ and $\mathbf{i}_{\tau:t-1}$, so we can focus on
\begin{eqnarray} \label{eq:SMC:3a}
\pi (\mathbf{i}_{\tau:t-1}, \btheta | \mathbf{x}_{0:t})
& \propto & \pi (\mathbf{x}_t| \btheta, \mathbf{x}_{0:t-1}, \mathbf{i}_{\tau:t-1})  \times \pi (\mathbf{i}_{\tau:t-1}, \btheta | \mathbf{x}_{0:t-1}), \end{eqnarray}
and then sample $\mathbf{i}_t$ as required.

The procedure we adopt maintains the number of occult infectives, $\mathbf{i}_{\tau:t-1}^U$ present at time $t-1$. For an individual, $j$ say, which becomes notified at time $t$ with no infection time prior to time $t$, we simply choose an individual, $l$ say, from the set of occult infectives at time $t-1$ who remains an occult infective at time $t$, and make individual $j$ infectious in place of individual $l$. That is, if individual $l$ had become infected at time $i_l$, this becomes the time at which individual $j$ now becomes infected with individual $l$ now assumed to be susceptible through to time $t$. We repeat the process until the new $\mathbf{i}_{\tau:t-1}$ satisfies \eqref{eq:SMC:3}.

The new particle, $(\btheta, \mathbf{i}_{\tau:t-1}^\ast)$ say, obtained from the above adjustment process is no longer sampled from $\pi (\btheta, \mathbf{i}_{\tau:t-1} | \mathbf{x}_{0:t-1})$ but from a density $\pi^\ast (\btheta, \mathbf{i}_{\tau:t-1} | \mathbf{x}_{0:t})$ which satisfies
\begin{eqnarray} \label{eq:SMC:4}
\pi^\ast (\btheta, \mathbf{i}_{\tau:t-1}^\ast | \mathbf{x}_{0:t}) = \sum_{\mathbf{i}_{\tau:t-1} = \mathbf{a}} q (\mathbf{a} \rightarrow \mathbf{i}_{\tau:t-1}^\ast) \pi (\btheta, \mathbf{a} | \mathbf{x}_{0:t-1}),
\end{eqnarray}
where $q (\mathbf{a} \rightarrow \mathbf{i}_{\tau:t-1}^\ast)$ is the probability a set of infection times $\mathbf{a}$ is adjusted to $\mathbf{i}_{\tau:t-1}^\ast$. Note that $q (\mathbf{i}_{\tau:t-1}^\ast \rightarrow \mathbf{i}_{\tau:t-1}^\ast) =1$ and that $\pi^\ast( \cdot | \mathbf{x}_{0:t})$ explicitly highlights the dependence on $\mathbf{x}_t$ as this informs the adjustment. For a given $\mathbf{i}_{\tau:t-1}$ there will  typically be multiple adjustments possible which result in  $\mathbf{i}_{\tau:t-1}^\ast$ satisfying \eqref{eq:SMC:3}.

The probability $\mathbf{i}_{\tau:t-1}$ is adjusted to $\mathbf{i}_{\tau:t-1}^\ast$ depends upon a number of factors. Let $u_{t-1}$ denote the total number of occult cases at time $t-1$, $v_t$ denote the total number of new infections at time $t$ and $b (=b ( \mathbf{i}_{\tau:t-1}, \mathbf{i}_{\tau:t-1}^\ast))$ denote the total number of difference in the infection sets. Then there are a total of $\binom{u_{t-1} - (v_t - b)}{b}$ readjustments which can be performed. We consider two adjustment strategies:-
\begin{enumerate}
\item Simple random sampling, where each of the readjustments are equally likely.
\item Sample according to the probability that an occult infection will be removed, {\it eg.}~the hazard function. That is, to switch individual $j$ with an occult individual $l$ with probability proportional to  $h_Q (t -i_l) = g_Q(t-i_l)/\pz (Q\geq t - i_l)$.
\end{enumerate}
For both adjustment schemes we take a random permutation of the individuals who need to be adjusted (moved from $\mathcal{S}_{t-1}$ to $\mathcal{I}_{t-1}$) and then  in turn, using sampling without replacement, select an individual from the candidates to switch with by choosing either uniformly at random or according to the  $h_Q (\cdot)$ as appropriate. The second sampling scheme, whilst being more involved, looks to take account of $\pi (\mathbf{x}_t| \btheta, \mathbf{x}_{0:t-1}, \mathbf{i}_{\tau:t-1})$ in creating samples $(\btheta,  \mathbf{i}_{\tau:t-1}^\ast)$.

In general, we can not move beyond \eqref{eq:SMC:4} without extensive work in calculating $\pi (\btheta, \mathbf{a} | \mathbf{x}_{0:t})$. However, for homogeneously mixing epidemics progress can be made by noting that for any $\mathbf{a}$ such that $q (\mathbf{a} \rightarrow \mathbf{i}_{\tau:t-1}^\ast) \neq 0$, there is the same number of infections at each time point with each set of infections equally likely. This leads to
\begin{eqnarray} \label{eq:SMC:6}
\pi (\btheta, \mathbf{a} | \mathbf{x}_{0:t}) =
\pi (\btheta, \mathbf{i}_{\tau:t-1}^\ast | \mathbf{x}_{0:t}), \end{eqnarray}
and thus, \eqref{eq:SMC:4} can be simplified to
\begin{eqnarray} \label{eq:SMC:7}
\pi^\ast (\btheta, \mathbf{i}_{\tau:t-1}^\ast | \mathbf{x}_{0:t}) &=&  \pi (\btheta, \mathbf{i}_{\tau:t-1}^\ast | \mathbf{x}_{0:t-1}) \sum_{\mathbf{i}_{\tau:t-1} = \mathbf{a}} q (\mathbf{a} \rightarrow \mathbf{i}_{\tau:t-1}^\ast).
\end{eqnarray}

For the simple random sampling, we can show that
\begin{eqnarray} \label{eq:SMC:10}
\pi^\ast (\btheta, \mathbf{i}_{\tau:t-1}^\ast | \mathbf{x}_{0:t})
&\propto & \binom{u_{t-1}}{v_t}^{-1} \pi (\btheta, \mathbf{i}_{\tau:t-1}^\ast | \mathbf{x}_{0:t-1}). \end{eqnarray}
(See the Supplementary Materials, Section 1.1 for details.) Consequently, on adapting a particle  $(\btheta, \mathbf{i}_{\tau:t-1})$ to $(\btheta, \mathbf{i}_{\tau:t-1}^\ast)$, we multiply the weight $w_{t-1}^i$ by $\binom{u_{t-1}}{v_t}$ for $(\btheta, \mathbf{i}_{\tau:t-1}^\ast)$ to better represent a sample from $\pi (\btheta, \mathbf{i}_{\tau:t-1}^\ast | \mathbf{x}_{0:t})$.

For the second sampling scheme using $h_Q(\cdot)$, we can show, see Supplementary Material, Section 1.2 for details, that \begin{eqnarray} \label{eq:SMC:11} \pi (\btheta, \mathbf{i}_{\tau:t-1}^\ast | \mathbf{x}_{0:t-1}) q (\mathbf{a}_{v_t} \rightarrow \mathbf{i}_{\tau:t-1}^\ast |\bomega^\prime), \end{eqnarray} provides an approximate estimate of $\pi^\ast (\btheta, \mathbf{i}_{\tau:t-1}^\ast |\mathbf{x}_{0:t})$, where $\bomega^\prime$ represents a random permutation of the $v_t$ notification times at time $t$ to be assigned and $q (\mathbf{a}_{v_t} \rightarrow \mathbf{i}_{\tau:t-1}^\ast |\bomega^\prime)$ is the probability of the adjustment given the ordering $\bomega^\prime$. Hence we take the adjustment weight to be  $q (\mathbf{a}_{v_t} \rightarrow \mathbf{i}_{\tau:t-1}^\ast |\bomega^\prime)^{-1}$.

\subsubsection{Propagating the Particles} \label{sss:propagate}

Let $\{ (\tilde{\btheta}_t^i, \tilde{\mathbf{i}}_{\tau:t-1}^i ); i=1,2,\ldots,N\}$ denote the particles generated by the adjustment step above at time $t$. Let $\tilde{w}_t^i$ denote the corresponding weight for the particle which is given by
\begin{eqnarray} \label{eq:SMC:12}
\tilde{w}_t^i &=& A_t  \times \pi ( \mathbf{x}_{0:t-1}, \mathbf{i}_{\tau:t-1} = \tilde{\mathbf{i}}_{\tau:t-1}^i  | \tilde{\btheta}_t^i) , \end{eqnarray}
 where $A_t$ denotes the adjustment weight from adjusting the particle to be consistent with the observed data derived in Section \ref{sss:consistent}.
Thus the weight consists of the adjustment derived in Section \ref{sss:consistent}, the prior on $\btheta$, the likelihood of the data (observed and augmented) up to the end of time $t-1$ and the probability of observing the given notifications on day $t$.
Given the discussion in Section \ref{sss:consistent}, $\tilde{w}_t^i$ does not reflect the true posterior weight (up to a constant of proportionality) of $\pi(\tilde{\btheta}_t^i, \tilde{\mathbf{i}}_{\tau:t-1}^i | \mathbf{x}_{0:t})$ but should give a good approximation of this quantity.

%
%
%

For $i=1,2,\ldots, N$, we draw $(\btheta_t^i, \mathbf{i}_{\tau:t-1}^i)$ from the  $\{ (\tilde{\btheta}_t^i, \tilde{\mathbf{i}}_{\tau:t-1}^i ); i=1,2,\ldots,N\}$ with the probability that  $(\btheta_t^i, \mathbf{i}_{\tau:t-1}^i) =  (\tilde{\btheta}_t^l, \tilde{\mathbf{i}}_{\tau:t-1}^l )$ given by $w_t^l /\sum_{k=1}^N  w_t^k$. We then complete the augmented data $\mathbf{y}_{\tau:t}^i$ required for the MCMC step by simulating the set of new (occult) infections from $\pi (\mathbf{i}_t | \mathbf{x}_{0:t-1}, \mathbf{i}_{\tau:t-1}^i, \btheta^i_t)$ and the future notification times of the occult infections by drawing an infectious period, $Q$, for each of the occult individuals, subject to the constraint that the resulting notification time is after time $t$.

%
%
%
%

\subsubsection{MCMC jittering of Particles} \label{sss:jitter}

For each particle $i$ we run the MCMC algorithm outlined in Section \ref{ss:MCMC} for $n_p$ steps to jitter the particles ready to take forward the final value of $(\btheta_t^i, \mathbf{i}_{\tau:t}^i)$ from the MCMC runs to the next time point. Supposing that the $\{(\btheta_{t-1}^i, \mathbf{i}_{\tau:t-1}^i); i=1,2,\ldots,N \}$ generated at time point $t-1$ represent draws from $\pi (\cdot | \mathbf{x}_{0:t-1})$, the particle generation process should ensure that the starting particles for each of the MCMC algorithm are approximately drawn from $\pi (\cdot | \mathbf{x}_{0:t})$. Furthermore, we do not expect the marginal density $(\btheta_{t-1}^i, \mathbf{i}_{\tau:t-1}^i)$ to change significantly in the move from $\pi (\cdot | \mathbf{x}_{0:t-1})$ to $\pi (\cdot | \mathbf{x}_{0:t})$. Therefore for the tuning of the random walk proposal variance $\Sigma_\lambda$ we use \eqref{eq:Sigma:1}, where $\Sigma_B$ is estimated using the empirical covariance matrix at time point $t-1$. For the number of infection times, $m$ and $m_U$, to update, we monitor the number of proposed moves accepted changes as required at the start of time point $t$ in line with Section \ref{ss:MCMC}.

Given that we expect the posterior distribution of the parameters to not differ significantly from time point $t-1$ to time point $t$, it should be hoped that moderate $n_p$ is sufficient, so that at the end of $n_p$ iterations of the MCMC runs the resulting $\{(\btheta_t^i, \mathbf{i}_{\tau:t}^i); i=1,2,\ldots,N \}$ form an approximate sample from $\pi (\cdot | \mathbf{x}_{0:t})$. In contrast to a standard MCMC algorithm where we run one chain for a large number of iterations to generate a sequence of dependent samples from the posterior distribution, we are using an MCMC kernel $n_p$ times on $N$ particles from an approximation of the posterior distribution to obtain a representative sample from the posterior distribution of interest. It is thus properties of the sample rather than an individual particle that are important to us, and it is not a concern that a particle has not {\it forgotten} its starting value at the end of the MCMC run. We explore this briefly in Section \ref{S:Theory} which gives a more theoretical treatment of the MCMC-within-SMC algorithm.

\section{MCMC-within-SMC Theory} \label{S:Theory}

In this Section we present a brief summary of how the MCMC-within-SMC works from a theoretical perspective.

The main aim of the MCMC-within-SMC algorithm is to obtain samples from $\pi (\btheta, \mathbf{y}_{\tau:t} | \mathbf{x}_{0:t})$. The samples from the MCMC-within-SMC will often be evaluated through the computation of expectations of functions of the parameters. That is,
for $t \in \mathbb{N}$ and a function $\phi (\cdot)$, we are interested in estimating
\begin{eqnarray} \label{eq:Theory:1} \mu_\phi^t &=& \ez [ \phi (\btheta_t)] \nonumber \\
&=& \int \phi (\btheta) \pi (\btheta | \mathbf{x}_{0:t}) \, d \btheta. \end{eqnarray}

The SMC algorithm is constructed to provide an (approximate) sample \begin{eqnarray} \label{eq:Theory:2} \{(\btheta_{t-1}^i, \mathbf{y}_{\tau:t-1}^i); i=1,2,\ldots,N\} \end{eqnarray}
from $\pi (\btheta, \mathbf{y}_{\tau:t-1} | \mathbf{x}_{0:t-1})$. A consistent estimator of $\mu_\phi^t$ using the samples from \eqref{eq:Theory:2} is provided by
\begin{eqnarray} \label{eq:Theory:3} \tilde{\mu}_\phi^t &=& \frac{\sum_{i=1}^N  \phi (\btheta_{t-1}^i) \times \pi (x_t, y_t | \btheta, \mathbf{y}_{\tau:t-1}, \mathbf{x}_{0:t-1})}{\sum_{i=1}^N    \pi (x_t, y_t | \btheta, \mathbf{y}_{\tau:t-1}, \mathbf{x}_{0:t-1})} .   \end{eqnarray}
As noted earlier $\pi (x_t, y_t | \btheta, \mathbf{y}_{\tau:t-1}, \mathbf{x}_{0:t-1})$ will often be equal to 0 with $x_t$ not being consistent with $\mathbf{y}_{\tau:t-1}$ leading to the estimator $\tilde{\mu}_\phi^t$ having a large variance. The adjustment of the particles (sample) in \eqref{eq:Theory:2} leads to
\begin{eqnarray} \label{eq:Theory:4} \{(\btheta_t^{i,0}, \mathbf{y}_{\tau:t-1}^{i,0}); i=1,2,\ldots,N\}, \end{eqnarray}
from $\pi^\ast ( \btheta, \mathbf{y}_{\tau:t-1}| \mathbf{x}_{0:t})$, where $\btheta_t^{i,0} = \btheta_{t-1}^i$. However, since we do not know $\pi^\ast ( \btheta, \mathbf{y}_{\tau:t-1}| \mathbf{x}_{0:t})$, even up to a constant of proportionality, it is not possible to construct an estimator along the lines of \eqref{eq:Theory:3}.

The MCMC runs for each particle generate  $\{(\btheta_t^{i,k}, \mathbf{y}_{\tau:t}^{i,k}); i=1,2,\ldots,N, k=0,1, \ldots,n_p \}$ with
\begin{eqnarray} \label{eq:Theory:5}
\hat{\mu}_\phi^{t,k} = \frac{1}{N} \sum_{i=1}^N \phi (\btheta_t^{i,k})
 \end{eqnarray}
providing a natural estimator for $\mu_\phi^t$ given in \eqref{eq:Theory:1}. The question is, how does the estimator given in \eqref{eq:Theory:5} varies with $k$? We can consider the mean square error (MSE) of $\hat{\mu}_\phi^{t,k}$ given by
\begin{eqnarray} \label{eq:Theory:6}
\ez \left[(\hat{\mu}_\phi^{t,k} - \mu_\phi^t)^2 \right] &=& var \left( \hat{\mu}_\phi^{t,k} \right) + \left\{ \ez [ \hat{\mu}_\phi^{t,k} ] - \mu_\phi^t \right\}^2. \end{eqnarray}
For $k=0$, we obtain an unbiased estimate of
\begin{eqnarray} \label{eq:Theory:7} \mu_\phi^{t,\ast} &=& \int \phi (\btheta) \pi^\ast (\btheta | \mathbf{x}_{0:t}) \, d \btheta. \end{eqnarray}
However, we are unable to quantify the difference between $\pi^\ast (\btheta | \mathbf{x}_{0:t})$ and $\pi (\btheta | \mathbf{x}_{0:t})$ and without the MCMC jittering of particles we have particle degeneracy as $t$ increases. On the other hand, as $k \rightarrow \infty$, we obtain  independent samples from the posterior distribution at time $t$ with $\ez [ \hat{\mu}_\phi^{t,k} ] \rightarrow \mu_\phi^t$.
Therefore
\begin{eqnarray} \label{eq:Theory:8}
var \left( \hat{\mu}_\phi^{t,k} \right) \rightarrow \frac{1}{N} var (\phi (\check{\btheta}_t) ), \hspace{1cm} \mbox{as } n \rightarrow \infty, \end{eqnarray} where $\check{\btheta}_t$ denotes the posterior distribution of $\btheta$ at timepoint $t$. Thus the MSE converges to $var (\phi (\check{\btheta}_t) )/N$ as $k \rightarrow \infty$, which corresponds to the MSE obtained from taking $N$ independent samples from $\check{\btheta}_t$.
Whilst, we can't evaluate \eqref{eq:Theory:6}, we observe that it is best to use $\hat{\mu}_\phi^{t,n_p}$, the final values of the MCMC run. This is because it reduces the dependence between the different MCMC runs caused by some particles having the same starting values of $\theta$ due to resampling at iteration $t-1$ and to reduce the effects of  $\{(\btheta_t^{i,0}, \mathbf{y}_{\tau:t-1}^{i,0}); i=1,2,\ldots,N\}$ being drawn from $\pi^\ast ( \btheta, \mathbf{y}_{\tau:t-1}| \mathbf{x}_{0:t})$.

This leaves the question, how large should $n_p$ be to balance convergence to the posterior distribution with computational requirements? In this paper we consider this through comparing, for different choices of $n_p$, the samples from the posterior distribution obtained using the MCMC-within-SMC with output from an MCMC algorithm. In the simulation studies and the FMD outbreak in Sections \ref{S:Simulation} and \ref{S:FMD}, respectively, we find $n_p=25$ to $n_p=500$ suffices to provide a good approximation of the posterior distribution depending on the size of the epidemic. Monitoring how $\hat{\mu}_\phi^{t,k}$ evolves with $k$ can be useful in determining whether $n_p$ needs to be made larger or a smaller value of $n_p$ will suffice. Note that we can vary $n_p$ at each time point $t$ and thus this can be assessed and updated as the algorithm is run.

\section{Simulation study} \label{S:Simulation}

In this Section we present a simulation study to gain a better understanding of how the SMC algorithm presented in Section \ref{S:Algorithm} performs. The simulation study is designed to address the key question, how does it compare with the {\it gold standard} of MCMC. In order to answer this question we compare the posterior distribution obtained using SMC with MCMC alongside studying how the SMC algorithm performs over multiple time points to see if we observe deterioration in the performance of the algorithm and how fast SMC is in comparison with MCMC. Further analysis of the simulation study is presented in the Supplementary Material, Section 2.


Two simulated data sets are considered one for an $SIR$ epidemic model and the other for an $SINR$ epidemic model. The simulations are spatial epidemic models with individuals located uniformly at random over the unit square. The probability that individual $k$ makes an infectious contact with individual $l$ on a given day is given by
\begin{eqnarray} \label{eq:sim:1} p_{kl} = (1-p) \exp(-\gamma d(k,l)), \end{eqnarray}
where $d(k,l)$ denotes the Euclidean distance between individuals $k$ and $l$. The parameter $\phi$ denotes the reduced infectivity level of notified farms. The infectious periods are independent and identically distributed according to $Q \sim {\rm Po} (a) +1$ and for the $SINR$ the length of the period from notification to removal is of fixed length $d$. Finally, letting $N_{pop}$ denote the population size we have the following parameter set.
\begin{table}[htpb!]
\begin{center}
\begin{tabular}{r c c c c c c c} \hline
 & $N_{pop}$ & $1-p$ & $\gamma$ & $\phi$ & $a$ & $d$ & Population Distribution \\ \hline
 $SIR$ Simulation & 500 & 0.025 & 15 & - & 3 & -& $U(0,1) \times U(0,1)$\\
 $SINR$ Simulation & 300 & 0.015 & 10 & 0.2 & 4 & 4& $U(0,1) \times U(0,1)$ \\ \hline
\end{tabular}
\caption{The settings used to generate the $SIR$ and the $SINR$ epidemics.}
\label{tab:SimSettings}
\end{center}
\end{table}

For both data sets the initial analysis of the data was on day $T=3$, three days after the epidemic is first observed with 6 removals and 5 notifications, respectively, for the $SIR$ and $SINR$ data sets. The epidemics are observed and analysed up until the end of the epidemics which are days 79 and 105 with 146 and 103 individuals infected, respectively, for the $SIR$ and $SINR$ data sets. Thus both simulations have approximately a third of the population infected.

We focus primarily on the estimation of the infection parameters $(p, \gamma, \phi)$ and the number of occult infections $u_t$ under the assumption that $a$ is known. This assumption is not unreasonable as the distribution of the infectious period for many diseases is well known. In the supplementary material we present a further $SIR$ simulation with a larger variance ($a=7$) on the infectious period distribution. We observe that $a$ is sensitive to the choice of prior but that the estimation of $\gamma$ is robust to miss-specification of $a$ with $p$ adapting such that $(1-p) \times (a+1)$ (probability of infection per day  times mean infectious period) is estimated well.

For the probabilities $p$ and $\phi$, $U(0,1)$ priors are chosen whilst for $\gamma$, ${\rm Gamma} (1.69,0.13)$ and ${\rm Gamma} (2.25,0.25)$ priors are chosen for the $SIR$ and $SINR$ epidemics, respectively. This corresponds to prior means (standard deviations) of $13$ $(10)$ and $9$ $(6)$ for $\gamma$ for the $SIR$ and $SINR$ epidemics, respectively.

For estimation of the parameters we used the SMC algorithm with $1000$ particles. The SMC algorithms were initiated at time $T=3$ with particles drawn from every $50$ iteration of the MCMC algorithm after a burn-in of $10,000$ iterations (total length of the MCMC run $60,000$ iterations). The SMC algorithm was then applied to each time point to update the posterior distribution of $(p, \gamma, \phi, u_t)$ with MCMC runs of length $n_p=25$ and $n_p=50$. For comparison the MCMC algorithm was run for $60,000$ iterations (10,000 iterations as burn-in) at every 5 time points.

The simulation study showed very good agreement between the estimates (posterior means and standard deviations) of the parameters for the SMC and MCMC algorithms throughout the course of the epidemic. In all cases as the epidemic progressed we obtained improved estimates of the infection parameters and good estimation of the number of occult cases $u_t$ which is crucial in being able to determine successful control measures.

We observed that whilst the SMC algorithm requires more computing resource than running the MCMC algorithm at every 5 time points, its embarrassingly parallel nature meant that the time consuming particle updates can be split into $P$ jobs to share across $P$ processors. We found that for $n_p =25$ and $n_p=50$ using $P=5$ and $P=10$, respectively, made the SMC algorithm faster than the MCMC algorithm and increasing $P$ had a substantial, close to linear, reduction in the time taken.

One concern with SMC methods is particle degeneracy. We observe that the number of unique particles sampled at each time point remains fairly constant throughout the course of the epidemic. The number of unique points drops when there are a larger number of removals ($SIR$) or notifications ($SINR$) on a given day. Throughout both simulations for $n_p=50$, the total number of unique particles at each time point remains above 100 and is between 250 and 600 for the majority of time points.

\section{2001 Foot-and-Mouth disease (FMD) outbreak} \label{S:FMD}

In this Section we consider the 2001 FMD outbreak in Cumbria. As noted in Section \ref{ss:FMD:desc}, Cumbria was the worst hit county in the outbreak accounting for over $40\%$ of all cases. A detailed summary of the data are presented in the Supplementary Material, Section 3.

\subsection{FMD model}

The temporal pattern of the FMD outbreak shows a very clear spatial spread leading to the incorporation of a distance kernel in the model. We follow \cite{Jewell09} in using the Euclidean distance between farms, due to both its simplicity and the work of \cite{Savill06} which has shown that Euclidean distance is a better predictor of transmission risk than the shortest and quickest routes via road, except where major geographical
features intervene.

Given the spatial spread of FMD, we use a discrete time version of the Cambridge-Edinburgh model (\cite{keeling2005models}) with attention focussed on the spread of FMD amongst farms that contain cattle and sheep as these species were the primary carriers of FMD, see \cite{keeling2001dynamics} and \cite{Jewell09}. Let $p_{kl}$ denote the probability that an infectious farm $k$ will make an infectious contact with a susceptible farm $l$ on a given day. We take $p_{kl}$ to be
\begin{eqnarray} \label{eq:FMD:1} p_{kl} = 1 - \exp \left\{ - \beta_0 (s_k + \beta_1 c_k)^{\chi_1} (s_l + \beta_2 c_l)^{\chi_2} \exp (- \gamma d (k,l))\right\}, \end{eqnarray}
where  $d(x,y)$ is the Euclidean distance between farms $x$ and $y$ and $s_x$ and $c_x$ are the total number of sheep and cattle, respectively, on farm $x$. We can view
\begin{eqnarray} \label{eq:FMD:2} \beta_0 (s_k + \beta_1 c_k)^{\chi_1} (s_l + \beta_2 c_l)^{\chi_2} \exp (- \gamma d (k,l)) \end{eqnarray} as the transmission rate between farms $k$ and $l$ and as such compare our transmission model with \cite{Jewell09}, (10) and the Cambridge-Edinburgh model presented in \cite{keeling2005models}. Note that $\beta_0$ denotes the baseline infection rate between sheep with $\beta_1$ and $\beta_2$ representing the relative infectivity and susceptibility, respectively, of cattle to sheep. These are in agreement with \cite{Jewell09} and \cite{keeling2005models}. The parameters $\chi_1$ and $\chi_2$ represent how infectivity and susceptibility, respectively, of a farm scale with size. In \cite{keeling2005models}, implicitly $\chi_1 = \chi_2 =1$, representing linear growth in infectivity and susceptibility, whilst \cite{Jewell09} replace  $(s_k + \beta_1 c_k)^{\chi_1}$ and $(s_l + \beta_2 c_k)^{\chi_2}$ by $(s_k^\chi + \beta_1 c_k^\chi)$ and $(s_l^\chi + \beta_2 c_k^\chi)$, respectively. We contend that it is more natural for the scaling factor $\chi$ to act on the overall size of the farm rather than the number of sheep and cattle separately. Also the results below support $\chi_1 \neq \chi_2$, that is, the size of the farm affects infectivity and susceptibility differently. Finally, we use an exponential distance kernel as opposed to the heavy-tailed kernel of \cite{Jewell09}.

We take the infectious periods to be ${\rm Po} (a) +1$ as in Section \ref{S:Simulation} with $a=5$ producing an infectious period with mean 6 and variance 5. With the addition of the notification period, this led to the mean total time from a farm being infected to being culled to be approximately $7.25$ days.

\subsection{Algorithm settings}

We focus our attention on the first 32 days of the FMD outbreak in Cumbria to simulate applying the MCMC-within-SMC to an emerging disease outbreak. The initialisation of the SMC algorithm took place at time $T=4$ at which point there were $m^N_4=15$ notifications and $m^R_4=6$ removals. We then ran the SMC algorithm forward $28$ days to time $t=32$ just after the peak of the epidemic when  $m_{32}^N=386$ and $m^R_{32}=361$.

For all parameters, except $\kappa$ for which we used a $U(0,1)$ prior, we choose an exponential prior with prior means of $10^{-3}$, $10^{-4}$, $1$, $1$, $0.5$ and $0.5$ for $\gamma$, $\beta_0$, $\beta_1$, $\beta_2$, $\chi_1$ and $\chi_2$, respectively. The low mean on the prior on $\gamma$ reflects that the distances between farms were in metres with the transformation to kilometres corresponding to a prior mean of 1 on the spatial effect.

To generate the initial particles for the SMC we run the MCMC algorithm for 50,000 iterations after a burn-in of 10,000 iterations taking the output from every $50^{th}$ iteration of the MCMC as an initial particle. For the MCMC-within-SMC we used $n_p=200$ and $n_p=500$ to test the affect of varying $n_p$. The FMD data set being larger than the simulated data sets necessitated taking $n_p$ to be larger for reasonable mixing of the particles. We considered both the adjustments proposed in Section \ref{sss:consistent} to make the particles consistent with the observed data.

In order to assess the performance of the SMC algorithm we ran the MCMC weekly at $t=11, 18, 25$ and $32$. Since the MCMC algorithm mixing becomes worse as the epidemic progresses so we used 1,000,000 iterations after a burn-in of 500,000 iterations to ensure convergence for comparisons.

\subsection{Results}

\begin{sidewaysfigure}[htbp!]
\centering
\includegraphics[width=\textwidth]{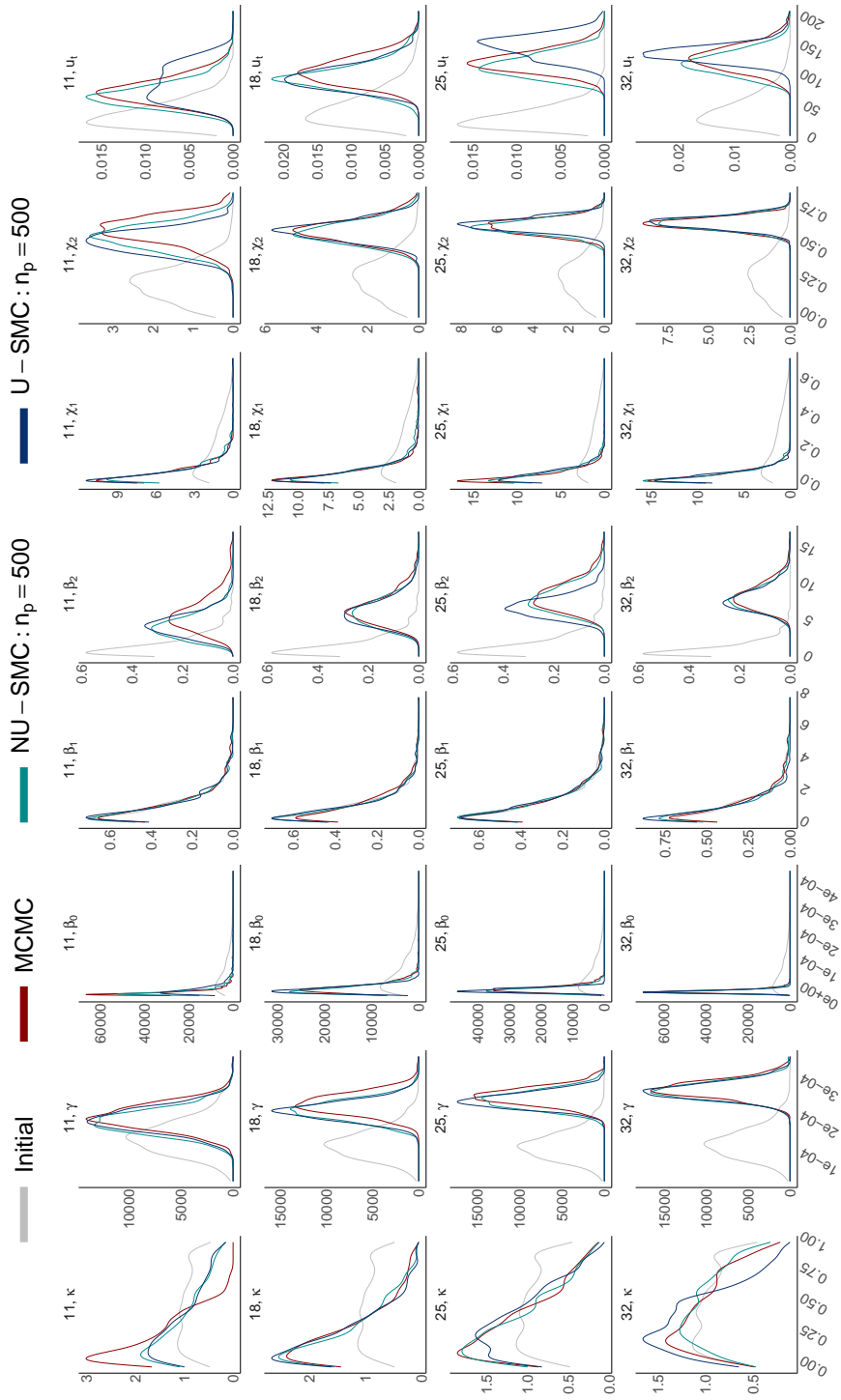}
\caption{A comparison of the densities generated using the outputs from the MCMC and SMC methods applied to the FMD data set. The SMC method has been applied using both uniform and non-uniform adjustment, both with $n_p=500$.}
\label{fig:FMD_Fixed_NonUniform500}
\end{sidewaysfigure}

We present analysis for the first 32 days with the results summarised in Table 2 and Figure \ref{fig:FMD_Fixed_NonUniform500} for days $t=11, 18, 25$ and $32$. The results presented in Table 2 are for the SMC algorithm with non-uniform adjustment of particles with similar results presented in Table 4 of the Supplementary Material for the uniform adjustment. The results show that there is good agreement between the SMC algorithm and MCMC algorithm estimates of the parameters with the non-uniform adjustment performing better. We observe, as expected, that the estimation of the parameters improves as $n_p$ increases. The uniform adjustment of particles with $n_p=200$ exhibited an over-estimation of the number of occult cases at time points $t=25$ and $t=32$ with a knock-on effect on some parameter estimates. This was corrected by either using a larger value of $n_p$ or the non-uniform adjustment.

\includegraphics[width=\textwidth]{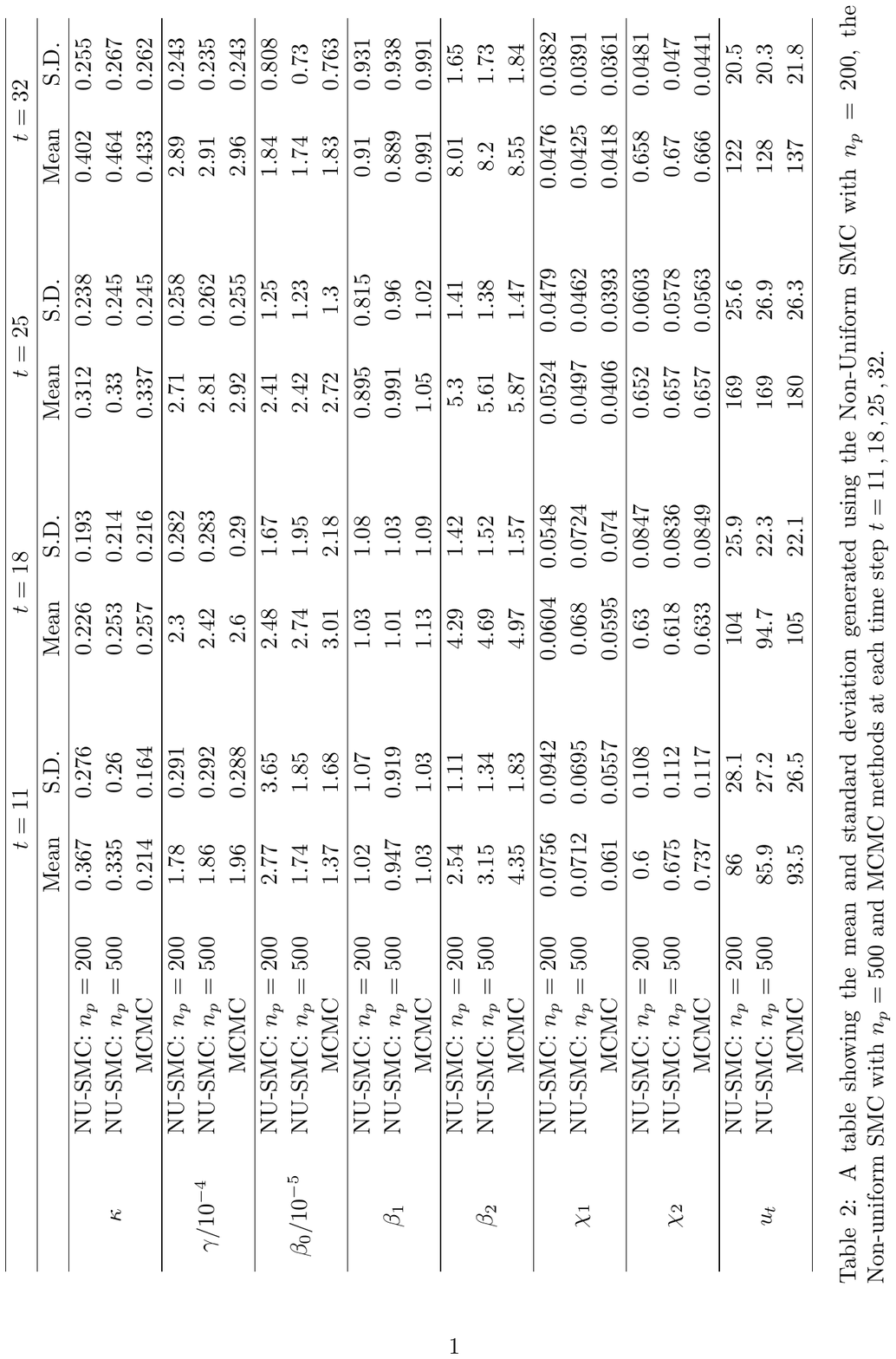}

The estimates of the parameters are informative about the spread of FMD. We observe that the estimation of $\beta_1$ (relative infectivity of cattle) closely mimics its prior. This is due to $\chi_1$ being estimated close to 0 throughout with the consequence that the value of $\beta_1$ has little impact on the likelihood. This suggests that the size of the farm has little impact on its infectivity, whilst with $\chi_2$ having a posterior mean around $0.666$ and posterior standard deviation of $0.044$ at time $t=32$, this suggests large farms are considerably more susceptible to becoming infected. We note that the posterior distribution supports $\beta_2 >1$, that is, cattle are more susceptible than sheep agreeing with previous findings, see \cite{Jewell09} and \cite{deardon2010inference}. We observe that posterior estimates of the parameters differ significantly from the initial estimates. The posterior means of $\chi_1$ and $\chi_2$ do not change significantly between $t=11$ and $t=32$ suggesting that the role of the size of the farm in the spread of the disease is not changing as the epidemic progresses, whereas other parameters, most notably $\gamma$, observe a marked change in the posterior mean. This suggests that some of the parameters could be varying with time to reflect changing behaviour in relation to the disease. The estimates of $\gamma$ increase as $t$ increases. This observation could indicate that control measures implemented during the outbreak increasingly prevented long range spread of FMD with a consequence that the posterior distribution increasingly supports local spread of FMD. Although we don't consider it in this paper, the SMC algorithm could easily be modified to allow for time-varying parameters to account for the evolution of disease dynamics, {\it ie.}~$\gamma$ changing with time.

\section{Conclusions} \label{S:Conc}

In this paper we have introduced an effective SMC scheme for analysing discretely observed epidemic processes. We have exploited, and in some cases developed, the efficient MCMC algorithms which exist for epidemic models to enable the SMC algorithm to update the particles in a timely manner. There are a number of interesting extensions of the work presented here.

Most of the research into epidemic models has focussed on continuous time models and it would be interesting to consider SMC algorithms for such models. The discrete time (daily) updates of the epidemic process and evaluation of the posterior distribution could be applied to a continuous epidemic model. Moreover, it should not be necessary to consider the SMC updates at regular observed intervals if so desired.

We have used data augmented MCMC to both seed the initial particles and within the SMC algorithm to update the particles. The initial MCMC is generally fast to run in an epidemic context where the time interval is short and only a few individuals have been infected. However, as the epidemic progresses with more infections over a longer time frame the MCMC updates take longer. In this paper we have taken the data augmentation updates over the whole of the epidemic process. An alternative would be to use a moving window of data of $K$ days, say, to be updated to reduce the slowing down of the algorithm. That is, at day $t$ take all augmented data prior to day $t- K$ to be fixed within the particle. The choice of a suitable $K$ to balance speed and mixing of the algorithm could be investigated.

Throughout this paper we have assumed that the parameters are constant through time. However, as mentioned in the FMD analysis in Section \ref{S:FMD} the SMC framework is perfectly suited to allowing for time varying parameters which enables the capturing of evolution of the disease or changes population dynamics in response to the disease outbreak.

We have assumed a given transmission kernel throughout this paper but it would be interesting to extend the SMC algorithm to select between competing transmission kernels. This could be done by starting with particles with a range of transmission kernels and studying which transmission kernel or kernels dominate the posterior distribution as the SMC algorithm progresses. 

\bibliographystyle{plain}

\end{document}